\documentclass[aps,prb,onecolumn,superscriptaddress,
showpacs,preprintnumbers,amsmath,amssymb
,floatfix
]{revtex4}
\usepackage{graphicx}
\usepackage{dcolumn}
\usepackage{bm}
\usepackage{amsmath}
\usepackage{amssymb}
\usepackage{revsymb}
 
\begin{document}

\affiliation{
Department of Physics and Astronomy, Georgia State
University, Atlanta, Georgia 30303, USA}
\affiliation{Chemistry Division, C-PCS, Los Alamos National Laboratory, 
Los Alamos, New Mexico 87545, USA}

\title{Nanoplasmonic Renormalization and Enhancement of Coulomb Interactions}
\affiliation{
Department of Physics and Astronomy, Georgia State
University, Atlanta, Georgia 30303, USA}

\author{Maxim Durach}
\affiliation{
Department of Physics and Astronomy, Georgia State
University, Atlanta, Georgia 30303, USA}
\author{Anastasia Rusina}
\affiliation{
Department of Physics and Astronomy, Georgia State
University, Atlanta, Georgia 30303, USA}
\author{Victor I. Klimov}
\affiliation{Chemistry Division, C-PCS, Los Alamos National Laboratory, 
Los Alamos, New Mexico 87545, USA}
\author{Mark I. Stockman}
\affiliation{
Department of Physics and Astronomy, Georgia State
University, Atlanta, Georgia 30303, USA}
\email{mstockman@gsu.edu}
\homepage{http://www.phy-astr.gsu.edu/stockman}

\date{\today}

\begin{abstract}
 
Nanostructured plasmonic metal systems are known to enhance greatly
variety of radiative and nonradiative optical
processes, both linear and nonlinear, which are due to the 
interaction of an electron in a molecule
or semiconductor with the enhanced local optical field of the surface
plasmons. Among them are surface enhanced Raman scattering (SERS)% 
\cite{Fleischmann_Hendra_McQuillan_Chem_Phys_Lett_26_163_1974_%
First_SERS_Observation,%
Jeanmire_Van_Duyne_J_Electronalytic_Chem_84_1_1977_SERS_Discovery,%
Albrecht_Creighton_JACS_99_5215_1977_SERS_Discovery,%
%Moskovits:1985_RMP_SERS,%
Kneipp:1997,%
Novotny_et_al_Science_2003_Fluorescence_and_SERS_from_Carbon_Nanotubes,%
Kneipp_Moskovits_Kneipp_SERS_Springer_Verlag_2005}, surface plasmon enhanced
fluorescence% 
\cite{Lakowicz_Anal_Biochem_2001_Review_Fluorescence,%
Shimizu_Woo_Fisher_Eisler_Bawendi:2002_PRL,%
Novotny_et_al_Science_2003_Fluorescence_and_SERS_from_Carbon_Nanotubes,%
Gerton_Wade_Lessard_Ma_Quake_Phys_Rev_Lett_93%
_180801_2004_ANSOM_10_nm_Resolution,%
Kuhn_Sandoghdar_et_al_PRL_2006_Fluorescence_Enhancement_and_Quenching,
Novotny_Stranick_ARPC_57_303_2006_NSOM},
fluorescence quenching in the proximity of metal surfaces,%
\cite{Dulkeith_et_al_Nano_Lett_2005_Fluorescence_Quenching,%
Lakowicz_Anal_Biochem_2001_Review_Fluorescence,%
Kuhn_Sandoghdar_et_al_PRL_2006_Fluorescence_Enhancement_and_Quenching}
coherent anti-Stokes Raman scattering (CARS)%
\cite{Ichimura_Hayazawa_Hashimoto_Inouye_Kawata_PRL_2004_Tip_Enhanced_CARS},
surface enhanced hyper-Raman scattering (SEHRS)%
\cite{Kneipp_et_al_PNAS_2006_Two_Photon_Hyper_SERS}, 
etc.
Principally different are numerous many-body phenomena that are due to the
Coulomb interaction between charged particles: carriers (electrons and holes)
and ions. These include carrier-carrier or carrier-ion scattering, 
energy and momentum transfer (including the drag effect),
thermal equilibration, exciton formation, impact ionization, Auger effects,%
\cite{Klimov_McBranch_PRL_1998_Auger_Recombination} etc. 
It is not widely recognized that these and other many-body effects can 
also be modified and enhanced by the surface-plasmon local fields. A special but 
extremely important class of such many-body phenomena is constituted by chemical
reactions at metal surfaces, including catalytic reactions. 
Here, we propose a general and powerful theory of the plasmonic enhancement 
of the many-body phenomena resulting in a closed expression 
for the surface plasmon-dressed Coulomb interaction. We illustrate this
theory by computing this dressed interaction explicitly for an
important example of metal-dielectric nanoshells,% 
\cite{West_Halas_Ann_Rev_Biomed_Eng_5_285_2003_%
Nanomaterials_sensing_therapeutics}
which exhibits a reach resonant behavior in both the magnitude and phase. 
This interaction is used to describe the nanoplasmonic-enhanced 
F\"orster energy transfer between nanocrystal quantum dots in the 
proximity of a plasmonic nanoshell. This is of great interest for 
plasmonic-enhanced solar cells and light-emitting devices.%
\cite{Klimov_et_all_Nature_429_642_2004_QW_QD_Fluorescence}
%Resonant plasmonic suppression of
%the Auger relaxation in nanocrystal quantum dots is another important 
%application.
Catalysis at nanostructured metal surfaces, nonlocal carrier scattering,
and surface-enhanced Raman scattering are discussed among other effects and
applications where the nanoplasmonic renormalization of the Coulomb
interaction may be of principal importance.

\end{abstract}

\pacs{%
78.67.-n, %
%       Optical properties of nanoscale materials and structures
%
%78.45.+h
%       Optical properties, condensed-matter spectroscopy and other
%       interactions of radiation and particles with condensed matter
%       Stimulated emission (see also 42.55 Lasers)
%42.50.-p
%       Quantum optics (for lasers, see 42.55 and 42.60)
%
%78.20.Bh
%       Optical properties of bulk materials and thin films
%       Theory, models, and numerical simulation
%
%42.55.-f
%       Lasers
%
%68.37.Uv,%
%       Near-field scanning microscopy and spectroscopy
%
%78.47.+p,
%       Time-resolved optical spectroscopies and other ultrafast
%       optical measurements in condensed matter
%
%42.50.Md,
%       Optical transient phenomena: quantum beats, photon echo,
%       free-induction decay, dephasings and revivals, and
%       optical 
%	nutation, and self-induced transparency
%
%42.65.Sf,
%       Dynamics of nonlinear optical systems; optical instabilities,
%       optical chaos, and complexity, and optical spatio-temporal
%       dynamics
%
71.45.Gm,
%       Electron structure:
%               Exchange, correlation, dielectric and magnetic
%       functions, plasmons
%
%42.65.Re,
%       Ultrafast processes; optical pulse generation and pulse compression
%
%05.45.+b,
%       Statistical physics and thermodynamics:
%       Theory and models of chaotic systems
%
%61.43.Hv,
%       Condensed matter, Structure of solids and liquids:
%       Disordered solids. Fractals, Macroscopic
%       Aggregates (including diffusion-limited aggregates).
%
%
%05.40.+j,
%       Statistical physics and thermodynamics:
%       Fluctuation phenomena, random processes, and Brownian
%       motion
%73.20 Fz,
%       Weak or Anderson localization. Surfaces, interfaces,
%       thin films and low-dimensional structures
%
73.20.Mf%
%       Collective excitations (including excitons,
%       polarons, plasmons and other charge-density excitations)
%
%85.35-p,
%       Nanoelectronic devices
%
%61.43.Hv,
%       Condensed matter, Structure of solids and liquids:
%       Disordered solids. Fractals, Macroscopic
%       Aggregates (including diffusion-limited aggregates).
%
%
%05.40.+j,
%       Statistical physics and thermodynamics:
%       Fluctuation phenomena, random processes, and Brownian
%       motion
}

\maketitle

Consider a system of charged particles situated in the vicinity of a
plasmonic metal nanosystem. For definiteness, we will assume that these particles
are electrons, although they can also be holes or ions of the lattice. One of the
examples of such systems is a semiconductor in the proximity of a
nanostructured metal surface. When an electron undergoes a transition
with some frequency $\omega$, this transition is accompanied by local
electric fields oscillating with the same frequency. These fields excite
surface plasmon (SP) modes with the corresponding frequencies whose
fields overlap in space with the transition fields. A property of these SP
eigenmodes is that they can be delocalized over the entire nanostructure
\cite{Stockman:2001_PRL_Localization}. The local optical fields of the
SPs can excite a resonant transition of
another electron. This process, which in the quantum-mechanical language
is the electron-electron interaction by the exchange of a SP quantum,
renormalizes (``dresses'') the direct interaction between these two
charges. As a result, the direct (``bare'') Coulomb interaction
between the electrons, $V(\mathbf r-\mathbf r^\prime)=
1/\left(\varepsilon_h \left|\mathbf r-\mathbf r^\prime\right|\right)$,
where $\varepsilon_h$ is the dielectric constant of the embedding medium, is
replaced by the dressed interaction $W(\mathbf r, \mathbf r^\prime;
\omega)$. This dressed interaction is generally not
translationally-invariant, i.e., it depends on coordinates $\mathbf r$
and $\mathbf r^\prime$ of both electrons; it also depends on
the transition frequency $\omega$. Note that $W$ is generally a complex
function, and its phase describes a delay inherent in the plasmonic reaction.
This phase varies sharply across plasmonic resonances, as we will show
later in this Letter.

\begin{figure}
\centering
\includegraphics[width=.8\textwidth]
{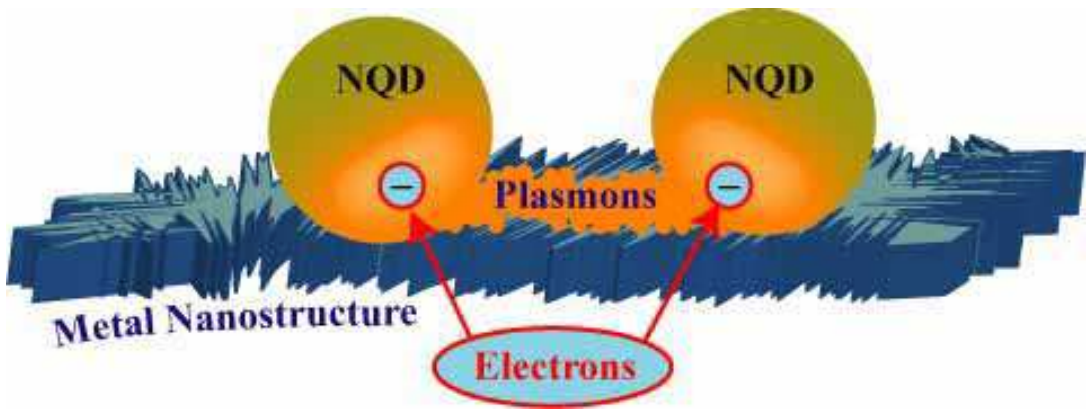}
\caption{\label{Schematic_3D.eps}
Schematic of metal nanostructure and semiconductor 
nanocrystal quantum dots (NQDs) situated in
its vicinity. The metal nanostructure is depicted by the dark blue
color. Two electrons are indicated in NQDs by the blue color and
the local plasmonic fields are schematically shown by the orange color.
}
\end{figure}

We schematically illustrate this SP-mediated electron-electron
interaction in Fig.\ \ref{Schematic_3D.eps} where we show electrons in
two semiconductor nancrystal quantum dots (NQDs) situated at the surface of a metal
nanostructure. The plasmonic fields, indicated by orange, excited by one
of the electrons interact with the other one in a different NQD. 
The interacting charges could also belong to the same quantum dot and could be
not only electrons but also holes.

The resonant nature of the electron-plasmon interaction may lead to the
significant enhancement of the dressed interaction $W$ with respect to
the bare one, $V$. A typical SP eigenmode tends to form ``hot spots" of
the local fields separated by distances on the order of the size
of the entire plasmonic nanostructure%
\cite{Stockman:2001_PRL_Localization}. Therefore one should expect that
the dressed interaction will also be delocalized over such distances,
i.e., be much more long-ranged than the bare Coulomb interaction.

%--------------------------------------------------------------------
%--------------------------------------------------------------------

The dressed interaction $W(\mathbf r, \mathbf r^\prime;
\omega)$, by definition, is the potential created at a point $\mathbf r$
by a charge positioned
at another point $\mathbf r^\prime$ and oscillating with frequency
$\omega$. We assume that the size of the system is
much smaller than any relevant electromagnetic length (radiation
wavelength, skin depth, etc.) and will use the quasistatic
approximation, which is conventional in nanoplasmonics.
In this case, $W$ satisfies the continuity equation
\begin{equation}
\frac{\partial}{\partial \mathbf r}\left[ \varepsilon(\mathbf r,\omega)
\frac{\partial}{\partial \mathbf r}
\,W(\mathbf r,\mathbf r^\prime; \omega)\right]=
-4\pi\delta (\mathbf r-\mathbf r^\prime)~,
\label{greenf_eqn}
\end{equation}
where dielectric function of the system $\varepsilon(\mathbf r,\omega)$
is expressed as 
$\varepsilon(\mathbf r,\omega)= 
\varepsilon_m(\omega)\Theta(\mathbf r) + \varepsilon_h [1-\Theta(\mathbf r)]$. 
Here, $\Theta(\bm{r})$ is the characteristic function
equal to 1 when $\mathbf r$ belongs to the metal and 0 otherwise, 
and $\varepsilon_m(\omega)$ is the dielectric function
of the uniform metal.

A general solution to this equation can be written in terms of the
retarded Green's function of the system $G^r$ as
\begin{equation}
W(\mathbf r,\mathbf r^\prime; \omega)=V(\mathbf r-\mathbf r^\prime)-
\int V(\mathbf r^{\prime\prime}-\mathbf r^\prime)
\frac{\partial^2}{\partial\mathbf r^{\prime\prime 2}}
G^r(\mathbf r,\mathbf r^{\prime\prime};\omega) d^3 r^{\prime\prime}~.
\label{gen_w}
\end{equation}
In Methods Section, we follow theory%
\cite{Stockman:2002_PRL_control,% 
Phys_Rev_B_69_054202_2004_Stockman_Bergman_Kobayashi_Coherent_Control} 
to outline derivation and properties of $G^r$,
which can be presented as a spectral
expansion over SP eigenmodes $\varphi_n(\mathbf r)$ and the
corresponding eigenvalues $s_n$ as
\begin{equation}
G^r(\mathbf r,\mathbf r^\prime;\omega)=
\sum_n \frac{s_n}{s(\omega)-s_n}\,
\varphi_n(\mathbf r)\varphi_n(\mathbf r^\prime)~,
\label{plasm_grf}
\end{equation}
where $s(\omega)=1/[1-\varepsilon_m(\omega)/\varepsilon_d]$ is the
spectral parameter. If the system is in an infinite space (or
the boundaries are remote enough), then
the use of Green's identity in Eq.\ (\ref{gen_w}) simplifies it to the
form 
\begin{equation}
W(\mathbf r,\mathbf r^\prime; \omega)=V(\mathbf r-\mathbf r^\prime)+
\frac{4\pi}{\varepsilon_h}\, G^r(\mathbf r,\mathbf r^\prime;\omega)~.
\label{W}
\end{equation}

This is a simple, yet, general and powerful result: the Coulomb
interaction is renormalized by the full retarded Green's function 
whose contraction also describes the nanoplasmonic enhancement of SERS%
\cite{Kneipp_Moskovits_Kneipp_SERS_Springer_Verlag_2005} and other
optical phenomena. The poles of Green's function (\ref{plasm_grf}) correspond
to the SP modes whose frequencies $\omega_n$ are given
by the equation $s(\omega_n)=s_n$. Close to such a frequency, $G^r$ becomes
large, proportional to the quality factor $Q_n$ of the SP resonance.%
\cite{Kneipp_Moskovits_Kneipp_SERS_Springer_Verlag_2005} This describes
the plasmonic renormalization and enhancement of the dressed Coulomb interaction.

%--------------------------------------------------------------------
\begin{figure}
\centering
\includegraphics[width=.80\textwidth]
{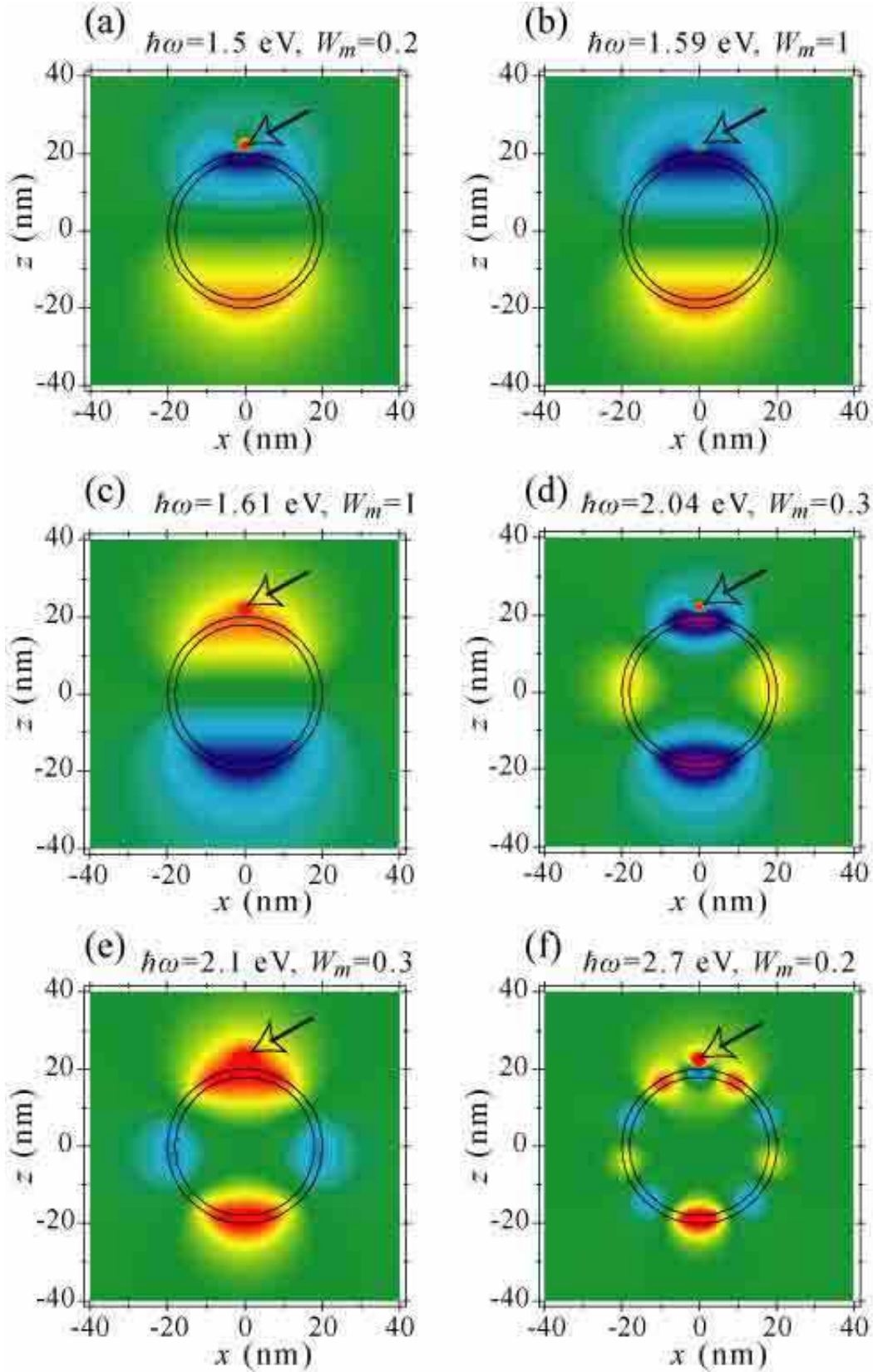}
\caption{\label{W.eps}
Renormalized (dressed) Coulomb interaction 
$\mathrm{Re}\,W(\mathbf r,\mathbf r^\prime;\omega)$
for silver nanoshell of aspect ratio $x=0.9$.
Point $\mathbf r^\prime$ is fixed and indicated by the black arrows in the upper
parts of the panels. The dependence of $\mathrm{Re}\,W(\mathrm r,\mathrm 
r^\prime)$ on $\mathbf r=\{x,z\} $ is shown by color coding in the panels for 
the cross section of the shell. The limits of this color coding are $\pm W_m$; 
the maximum $W_m$ and higher values of $\mathrm{Re}\,W$ 
are depicted by red, and the minimum $-W_m$ and 
lower values of $\mathrm{Re}\,W$ are shown by blue. Frequencies $\omega$ are 
indicated in the panels.
}
\end{figure}
%--------------------------------------------------------------------

We will illustrate the plasmonic renormalization and enhancement of the
Coulomb interaction using a metal nanoshell as a nanoplasmonic system. Such
nanoshells have significant fundamental and applied interest.%
\cite{West_Halas_Ann_Rev_Biomed_Eng_5_285_2003_%
Nanomaterials_sensing_therapeutics} For nanoshells, the renormalized
Coulomb interaction is derived in Methods as Eq.\ (\ref{W_shell}).
We depict the resonant behavior and renormalization (enhancement) of the
Coulomb interaction in Fig.\ \ref{W.eps} for a silver nanoshell 
with aspect ratio $x=0.9$ deposited on 
a dielectric core with permittivity $\varepsilon_h=2$ embedded into a
host with the same permittivity.  Silver dielectric function 
is adopted from the experimental data.\cite{Johnson:1972_Silver} 
For this specific 
nanoshell, the lowest dipole eigenmode (quantum numbers $L=1,P=-$) has 
eigenfrequency $\hbar\omega_{1-}=1.60$ eV. For a red-detuned  (from the 
dipole SP resonance) electronic transition frequency 
$\hbar\omega=1.5$ eV, the renormalized interaction is displayed in 
Fig.\ \ref{W.eps}(a). Very close to the singular point $\mathbf r=\mathbf 
r^\prime$, the renormalization is reduced to conventional dielectric 
screening: this is displayed by the opposite sign (blue color in the panel) of 
$\mathrm{Re}\,W(\mathbf r,\mathbf r^\prime;\omega)$ with respect to the bare 
Coulomb potential (the red dot pointed to by  the arrow). On the opposite 
side of the nanosphere, there is a ``mirror image'' of the dressed interaction 
potential $\mathrm{Re}\,W$ oscillating in phase (as indicated by the red color) 
with the initial field. This shows that the nanoplasmonic effects 
greatly extend the range of the dressed Coulomb interaction. 
This is due to the delocalization the dipole eigenmode, which defines 
the nanoplasmonic effects in this spectral region.

For $\hbar\omega=1.59$ eV, which is very close to (but still red-detuned 
from) the dipole SP resonance, the 
real part of the renormalized Coulomb interaction potential, 
$\mathrm{Re}\,W(\mathbf r,\mathbf r^\prime;\omega)$ is displayed in Fig.\ 
\ref{W.eps}(b). In this case, the dynamic screening of the bare potential 
becomes very strong, which is seen from the diminished magnitude and radius of 
the Coulomb potential shown as the small red dot pointed by the arrow.
The $W$ interaction, however, is strongly delocalized both around the point $\mathbf 
r^\prime$ where the charge is situated (manifested by the intense blue ``cloud'') 
and at the opposite pole of the nanoshell where the sign of the interaction is 
the same as that for the bare charge (red color).

The situation changes dramatically for a frequency $\hbar\omega=1.61$ eV, which 
is slightly above the dipole resonant frequency -- see Fig.\ \ref{W.eps}(c). In 
this case, the SP renormalization is actually dynamic {\it anti}-screening: 
both the radius and strength of the Coulomb interaction in the vicinity of the 
initial charge are increased. The renormalized potential oscillates in 
phase with the bare Coulomb potential, as indicated by the red color 
of the cloud around the arrow. There is also a 
very strong interaction potential on the diametrically opposite side of
the nanoshell, which oscillates out of phase (shown by blue). 
Thus, close to but slightly blue-detuned from the SP resonance, the 
dressed (renormalized) potential for $\mathbf r$ in the vicinity of
$\mathbf r^\prime$ becomes very large, which can 
be described as the nanoplasmonic enhancement due to the dynamic antiscreening. 

As frequency $\omega$ increases further [Fig.\ \ref{W.eps} (d)-(f)],
higher-multipole SP eigenmodes start to contribute to the Coulomb potential dressing, 
starting with the quadrupole in panels (d) and (e). In all the cases, the
screening in the vicinity of the charge for frequencies to 
the red of the resonance changes to the antiscreening for blue spectral 
detuning. The dressed potential is delocalized over the surface of the nanoshell, 
thus becoming extremely long-ranged. This is a general property of the nanoplasmonic 
renormalization of the Coulomb interaction: the range of the dressed
interaction always extends over the entire nanoplasmonic system. 
This effect is due to the absence of the strong localization of the SP 
eigenmodes, cf. Ref.\ \onlinecite{Stockman:2001_PRL_Localization}. 

In Fig.\ \ref{W.eps}, we have plotted only the real part of the renormalized 
Coulomb potential $W(\mathbf r,\mathbf r^\prime;\omega)$. However, its imaginary 
part (not shown) is also important. In the resonant cases, it is greatly 
enhanced and delocalized over the entire nanosystem. Because of the underlying 
$\pi/2$ phase shift, it does not interfere with the real part of $W$. 
It always increases the strength and contributes to the
delocalization of the dressed interaction.

One of the many-body effects that is affected by the nanoplasmonic renormalization of 
the Coulomb interaction is the F\"orster resonant energy transfer (FRET).% 
\cite{Foerster_Annalen_der_Physik_437_1_2_55_1948_%
Zwischenmolekulare_Energiewanderung_und_Fluoreszenz,%
Dexter_1953_Foerster_transfer} 
It has been proposed theoretically%  
\cite{Hua_Gersten_Nitzan_J_Chem_Phys_83_3650_1985_%
Theory_of_energy_transfer_between_molecules_near_solid_state_particles,%
Govorov_Lee_Kotov_PRB_plasmon-enhanced_Foerster_transfer_between_QD}
and observed experimentally% 
\cite{Lakowicz_An_Biochemistry_2002_Radiative_Decay_Engineering}
that the F\"orster transfer between chromophores in the proximity of a
metal spheroid is enhanced by the SP effect. Note that SP-mediated 
energy transfer across a metal film has also been observed.%
\cite{Andrew_Barnes_Science_306_1002_2004_%
Energy_Transfer_Across_Metal_Film_by_Surface_Plasmon_Polaritons} 
Below, as an illustration of our general theory,
we consider FRET for chromophores at the surface of a
metal/dielectric nanoshell. As we have already mentioned, the nanoshells
are spectrally tunable and can
have their SP eigenmodes shifted by frequency to the red and near-ir
spectral regions,\cite{West_Halas_Ann_Rev_Biomed_Eng_5_285_2003_%
Nanomaterials_sensing_therapeutics} which results in increased values
of the resonance quality factor due to lower dielectric losses of the
metal,\cite{Johnson:1972_Silver} and, hence, enhanced plasmonic effects.

The FRET is an electron-electron interaction (many-body) effect that is due to
dipole electronic transitions. It is described by Hamiltonian that is
a dipolar expansion of interaction (\ref{W}): 
\begin{equation}
H^\prime_{\rm FRET}=
\left(\mathbf d_d\frac{\partial}{\partial\mathbf r}\right)
\left(\mathbf d_a\frac{\partial}{\partial\mathbf r^\prime}\right)\,
W(\mathbf r,\mathbf r^\prime;\omega)~,
\label{H_FRET}
\end{equation}
where $\mathbf d_d$ and $\mathbf d_a$ are the dipole operators of the
two interacting electrons (the energy donor and acceptor)
at points $\mathbf r$ and $\mathbf r^\prime$, respectively.
Note that the SPs of all multipolarities are taken into account
by Eq.\ (\ref{H_FRET}).

For certainty, we will consider FRET between the electrons belonging
to two different NQDs, similarly to what is schematically 
illustrated in Fig.\ \ref{Schematic_3D.eps}. It is known that the direct
(without SP participation) FRET occurs between two NQDs only at
very short distances, on the order of just a few nanometers.%
\cite{Kagan_Murray_Nirmal_Bawendi_PRL_1996_NQD_Energy_Transfer,%
Crooker_Hollingsworth_Tretiak_Klimov_energy_transfer_in_quantum_dot_assemblies_2002,%
Achermann_Petruska_Crooker_Klimov_energy_transfer_in_quantum_dot_assemblies_2003} 
We will be interested in the FRET over larger distances where it occurs
predominantly via an SP-mediated process. 
We assume that the transitions $i\leftrightarrow f$ between the initial
and final states in NQDs are unpolarized, i.e., the corresponding
transition dipole moments $\left(\mathbf d_d\right)_{if}$ and
$\left(\mathbf d_a\right)_{if}$ are randomly-oriented vectors. In such a case,
substituting Hamiltonian (\ref{H_FRET}) into the Fermi Golden Rule
and averaging over the dipole orientations, we
obtain an expression for the plasmon-enhanced FRET rate $\gamma_F$,
\begin{equation}
\gamma_F=\frac{2\pi\left|\left(\mathbf d_d\right)_{if}\right|^2
\left|\left(\mathbf d_a\right)_{if}\right|^2}
{9\hbar^2} 
\left|W_{\alpha\beta}(\mathbf r,\mathbf r^\prime;\omega)\right|^2 J~,
\label{gamma_F}
\end{equation}
where $\alpha,\beta=x,y,z$ are vector indices (repeated indices imply
summation), $W_{\alpha\beta}$ is a dyadic renormalized interaction 
defined in Methods Section by Eq.\ (\ref{diadic_Greens}), and $J$ is the
spectral overlap integral. If the energy donor and acceptor transitions 
$i\leftrightarrow f$ both have Lorentzian shapes
with the same central frequency $\omega$ (i.e., are resonant) and
have homogeneous widths of $\gamma_d$ and $\gamma_a$, then 
$J=(2/\pi)\left(\gamma_a+\gamma_d\right)^{-1}$.

%--------------------------------------------------------------------
\begin{figure}
\centering
\includegraphics[width=.8\textwidth]
{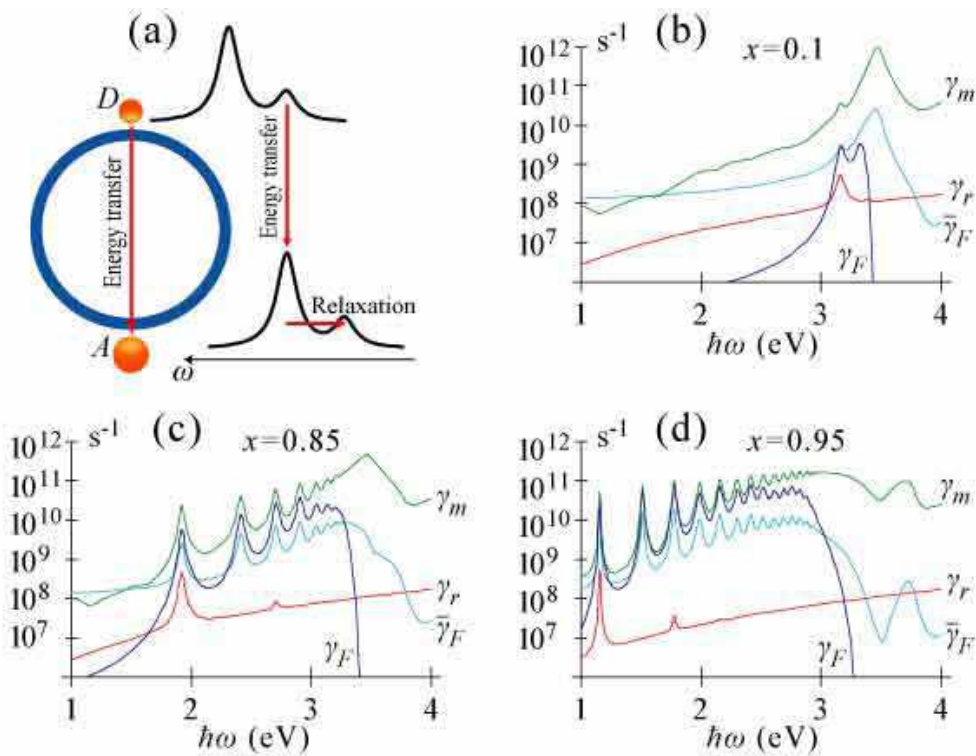}
\caption{\label{rates.eps}
Transfer and relaxation rates for NQDs at the outer surface of silver
nanoshell, modified and enhanced by SPs. (a) Schematic of the system and
energy transfer processes. A nanoshell is indicated by a blue circle, and
the donor and acceptor NQDs are labelled by D and A, correspondingly. The
frequency distributions of the transition oscillator strengths of  
the donor and acceptor NQDs are
shown by bold black curves. The energy transfer between NQDs and 
subsequent relaxation are indicated by red arrows. (b)-(c) The
nonradiative and radiative relaxation rates  (in the logarithmic scale)
for NQDs on nanoshells for the aspect ratios $x$ specified in the panels.
The FRET rate $\gamma_F$
(\ref{gamma_F}) for two NQDs situated on the opposite poles of a nanoshell 
[cf. panel (a)] is shown by the blue curves.
The SP-mediated FRET rate averaged over the position of the acceptor on
the nanoshell $\bar\gamma_F$ (\ref{bar_gamma_F}) is shown by the light-blue curves. 
The rate of
transfer to the metal $\gamma_m$ (\ref{bar_gamma_m}) is plotted by the green
curves. The radiative rate $\gamma_r$ (\ref{gamma_r}) for a NQD at the surface 
of the nanoshell is depicted by the red curves. 
}
\end{figure}
%--------------------------------------------------------------------

We compute the FRET rate for CdSe NQDs that are situated at the surface
of a silver nanoshell. For certainty, we further assume that they are
positioned at the opposite sides of the nanoshell, as illustrated in
Fig.\ \ref{rates.eps} (a). In CdSe NQDs, the lowest-energy, emitting
exciton state has a relatively weak oscillator strength and is shifted
by a few tens of meV (global Stokes shift) to the red with respect to
the stronger absorbing transition [see spectra shown by the bold black lines
in Fig.\ \ref{rates.eps}(a)]. Because of this significant Stokes shift,
the energy transfer occurs with an appreciable efficiency only between NQDs
with some size difference, which provides resonant coupling of the
emitting state of the donor (the smaller NQD) to the absorbing state of
the acceptor (the larger NQD).%
\cite{Crooker_Hollingsworth_Tretiak_Klimov_%
energy_transfer_in_quantum_dot_assemblies_2002,%
Achermann_Petruska_Crooker_Klimov_energy_%
transfer_in_quantum_dot_assemblies_2003}
This process is schematically illustrated in Fig.\ \ref{rates.eps}(a).

After the F\"orster energy transfer, indicated by the
vertical red arrow, there is a fast ($<1$ ps lifetime)
relaxation into the lower-frequency
state of the acceptor, which precludes the energy transfer back to the
donor and allows one to use the perturbation  theory result of Eq.\ 
(\ref{gamma_F}) describing an irreversible transfer. We adopted NQD
parameters for room temperature from Ref.\ 
\onlinecite{Crooker_Hollingsworth_Tretiak_Klimov_%
energy_transfer_in_quantum_dot_assemblies_2002}: $J=0.004$ cm
(or, $0.13$ ps); $d_a=25$ Debye, and $d_d=4.4$ Debye. Note that from this
value of $J$, one can obtain an estimate for the homogeneous width of
the donor and acceptor transitions $\gamma_a+\gamma_d=31$ meV, which
implies that these transitions are significantly broadened at room
temperature, probably, due to the electron-phonon interaction. 

For certainty, we
consider nanoshells with the outer radius $a=20$ nm. The donor and
acceptor NQDs are separated by 2 nm from the metal surface.
The FRET rates along with the rates of the nonradiative and radiative
decays of the NQDs are displayed in Fig.\ \ref{rates.eps} (b)-(d). We
show both the FRET rate $\gamma_F$ (\ref{gamma_F}) for the transfer
across the diamater of the nanoshell (dark-blue curves) and
the FRET rate $\bar\gamma_F$  (\ref{bar_gamma_F}) 
averaged over the random position of the
acceptor on the surface of the NQD 
(light-blue curves). This averaged rate is calculated and shown per a
single acceptor; if several acceptors participate in the energy
transfer, this rate should be multiplied by the
number of such acceptors on the nanoshell. Also shown by the green curves is
rate $\gamma_m$ (\ref{bar_gamma_m}) of nonradiative relaxation due to
the energy transfer to the metal. The rate the radiative decay $\gamma_r$
(\ref{gamma_r}) is depicted by the red curves.

For all aspect ratios [Fig.\ \ref{rates.eps} (b)-(d)], the FRET rates
$\gamma_F$ and $\bar\gamma_F$ are enhanced at the frequencies of the SP
eigenmodes, which is manifested by the corresponding resonant peaks in
the graphs. The number of the efficiently contributing multipoles
progressively increases from two (dipole and quadrupole) for a
low aspect ratio [$x=0.1$, panel (a)] to very many multipoles for high
aspect ratios [$x=0.85-0.95$, panels (c) and (d)]. With increasing 
aspect ratio $x$, the corresponding FRET rates in these peaks (at the
resonant SP frequencies) become higher by orders of magnitude.
Especially large this FRET transfer rate is in the lower-frequency
spectral range, which is due to a higher plasmonic
quality of the metal. 

The transfer to metal is an important process because it competes with
FRET. The nonradiative relaxation rate in the donor state of isolated
NQDs is low ($<0.05~\mathrm{ns}^{-1}$); 
therefore the energy transfer rate to metal $\gamma_m$ is the major
factor that determines the quantum yield of FRET: $Q_F=
\gamma_F/(\gamma_m+\gamma_F)$. We can see from Fig.\ \ref{rates.eps} (b)
that for low-aspect ratio nanoshells $\gamma_F\ll \gamma_m$ and, hence,
$Q_F\ll 1$ implying that the FRET is inefficient. The FRET of low efficiency 
was also found for solid nanospheres in Ref.\ 
\onlinecite{Govorov_Lee_Kotov_PRB_plasmon-%
enhanced_Foerster_transfer_between_QD}.
As the aspect ratio increases
[panels (c) and (d)], this metal-quenching rate $\gamma_m$ also shows
plasmonic peaks; however, it is enhanced significantly less than the
FRET rate. Correspondingly, for $x=0.95$ the FRET quantum yield is
rather high, $Q\sim 0.5$, in the red and near-ir regions of the spectrum. 
In contrast, the radiative rate $\gamma_r$ [Eq.\ (\ref{gamma_r})] is  orders of
magnitude lower than $\gamma_m$; correspondingly, the photoemission
quantum yield is very low. Note that the radiative rate is enhanced only 
in the odd-multipole (dipole, octupole,
etc.) SP resonances due to the parity selection rule. 

The overall conclusion from Fig.\ \ref{rates.eps} is that the SP
renormalization of the Coulomb interaction causes a strong enhancement
of the F\"orster transfer at large distances, across the entire nanoshell
and over a wide range of frequencies where many multipole SP resonances 
contribute. This enhancement is especially pronounced for high-aspect
ratio nanoshells in the red and near-ir frequency range where
rather high FRET efficiency is predicted.

To summarize briefly, significant renormalization 
of the Coulomb interaction between charged particles (electron, holes,
and lattice ions) in the vicinity of a plasmonic
nanosystem is demonstrated for transitions resonant with SP eigenmodes in
the nanosystem. There are three important features of this renormalized
interaction $W(\mathbf r, \mathbf r^\prime;\omega)$ that we have shown
above and would like to reemphasize here. (i) This renormalization
(enhancement) is highly resonant. Its phase depends on the frequency
detuning of the electronic transitions with respect to the SP resonant
frequency, changing by $\pi$ (from in-phase to out-of-phase and {\it vice
versa}) as frequency $\omega$ scans from the red to blue side of the SP resonance. 
(ii) The renormalized interaction $W(\mathbf r, \mathbf r^\prime;\omega)$ 
is long-ranged: the effective interaction length is on the order of 
the size of the entire nanosystem. (iii) This
renormalization and plasmonic enhancement of the Coulomb interaction is
a universal effect, which should affect a wide range of physical phenomena in the
vicinity of the metal nanoplasmonic systems: scattering between
charge carriers and the carriers and ions, ion-ion interactions,
exciton formation, etc.  One of the enhanced and long-range phenomena, which 
is due to the nanoplasmonic renormalization of the
Coulomb interaction, is the F\"orster energy transfer that becomes
effiecient for high-aspect ratio nanoshells.

Among other potentially very important applications of this theory are
chemical reactions and catalysis on nanostructured metal surfaces.
Chemical reactions occur due to the Coulomb interaction between charged particles
(electrons and ions) at small distances. In many cases, nanostructured
metals, in particular noble metal nanostructures with pronounced
plasmonic behavior, are good catalysts. The results of this theory
show that the Coulomb interaction at small distances is significantly
renormalized. As we have emphasized in the previous paragraph,
this renormalization is highly resonant depending on the frequency
$\omega$ of a transition that controls the chemical transformation 
(for example, breaking a chemical bond, establishing a desired new bond, 
isomerization, etc.). This renormalization leads to suppression
of the Coulomb interaction for the red detunings  from the
plasmonic resonance. In contrast, this
local interaction is greatly enhanced for blue detunings from the
SP resonance. This resonant effect opens up an avenue toward
``designer" nanostructured catalysts that can, e.g., favor one specific
reaction path over others.
%
%In addition to this, we would like here to briefly introduce
%important effects not discussed above, which may be subjects of the
%scrutiny in the near future.  The first of them is the possible suppression
%of the Auger relaxation. It is known that the Auger
%recombination and relaxation of excitons in NQDs is an important
%Coulomb-interaction phenomenon that leads to optical losses and limits
%the applications of NQDs as laser and nonlinear optical media.%
%\cite{Klimov_McBranch_PRL_1998_Auger_Recombination}  The
%nanoplasmonic renormalization of the Coulomb interaction, in principle,
%allows for a unique possibility to suppress this adverse effect. In
%fact, for small red detunings from the SP resonances, the nanoplasmonic
%renormalization at distances $|\mathbf r - \mathbf r^\prime|$ of a few
%nanometers, which are only relevant for the Auger effect, results in the
%strong suppression of the Coulomb interaction. This can be seen as
%manifested by the blue-green-color areas around the points $\mathbf
%r^\prime$ (indicated by arrows) in Fig.\ \ref{W.eps} (a), (b) and (d).
Another class of important effects, 
which are based on the nonlocal nature of the renormalized Coulomb 
interaction in the vicinity
of a nanoplasmonic system, is the nonlocal (cross) scattering. 
A charge at some point can undergo a transition, scattering from a charge at a
remote position, thus ``teleporting" momentum and energy. One
important instance of such a remote scattering is a
nonlocal SERS where an electronic transition occurs at one point 
but the vibrational energy is deposited at a distant point 
of the nanoplasmonic system.

\section[*]{Methods}
\label{Methods}

\subsection{Eigenproblem and Green's Functions of Nanoplasmonic System}

In this Section, for the sake of completeness and convenience, we
outline obtaining Green's function expressions within the framework
of spectral theory
\cite{Stockman:2001_PRL_Localization,%
Phys_Rev_B_69_054202_2004_Stockman_Bergman_Kobayashi_Coherent_Control,%
Kneipp_Moskovits_Kneipp_SERS_Springer_Verlag_2005}.
Consider a system consisting of a metal with dielectric permittivity
$\varepsilon_m(\omega)$, dependent on optical frequency $\omega$,
embedded in a dielectric background with dielectric
constant $\varepsilon_h$. The geometry of the system is described by the
characteristic function $\Theta(\mathbf r)$, which equals 1 in the metal
and 0 in the dielectric. Material properties of the system are described
by the spectral parameter $s(\omega)=\left[1-
\varepsilon_m(\omega)/\varepsilon_h\right]^{-1}$.

For a nanosystem, which has all sizes much smaller than the relevant
electrodynamic dimensions (radiation wavelength across the propagation
direction of the excitation wave and the skin depth in this direction), the
quasistatic approximation is applicable. In such a case, we define
retarded Green's function $\bar G^r({\bf r}, {\bf r}^\prime; \omega)$
as satisfying the continuity equation with the $\delta$-function right-hand side,
\begin{equation}
\left[\frac{\partial}{\partial {\bf r}}\Theta({\bf %
r})\frac{\partial}{\partial {\bf r}}%
-s(\omega)\frac{\partial^2 }{\partial {\bf r}^2}\right] 
\bar G^r({\bf r}, {\bf r}^\prime; \omega)
=
\delta({\bf r}-{\bf r}^\prime)~
\label{Greens_equation}
\end{equation}
and homogeneous Dirichlet or Dirichlet-Neumann boundary conditions.

It is convenient to expand the Green's function over eigenmodes
$\varphi_n(\mathbf r)$ and the corresponding eigenvalues $s_n$ that
satisfy a homogeneous counterpart of Eq.\ (\ref{Greens_equation}) 
\begin{equation}
\left[\frac{\partial}{\partial {\bf r}}\Theta({\bf %
r})\frac{\partial}{\partial {\bf r}}%
-s_n\frac{\partial^2 }{\partial {\bf r}^2}\right] 
\varphi_n({\bf r})=0~,
\label{phi_n}
\end{equation}
with the homogeneous boundary conditions. 
This spectral expansion of the Green's 
function can be readily obtained from Eq.\ (\ref{Greens_equation}). It
has an explicit form
\begin{equation}
\bar G^r({\bf r}, {\bf r}^\prime; \omega)=
\sum_n \frac{\varphi_n({\bf r})\,\varphi_n({\bf r}^\prime)}
{s(\omega)-s_n}~.
\label{Greens_expansion}
\end{equation}

Two features of this expansion are important. First, it separates the
dependencies on geometry and material properties. The geometrical
properties of the nanosystem enter only through the eigenfunctions
$\varphi_n$ and eigenvalues $s_n$ that are independent on the material
properties of the system. Therefore they can be computed for a given
geometry once and stored, which simplifies and accelerates further
computations. Complementary, the material properties of the system enter
Eq.\ (\ref{Greens_expansion}) only through a single function: spectral
parameter $s(\omega)$. 
The second important feature is that this Green's function satisfies
exact analytical properties due to the form of Eq.\
(\ref{Greens_expansion}) that contains only simple poles in the lower
half-plane of the complex frequency $\omega$ and does not have any
singularities in the upper half-plane of $\omega$. Consequently, $\bar
G^r$ is a retarded Green's function that automatically guarantees the
causality of the results of time-dependent calculations. Namely, the
Green's function in time domain satisfies the condition 
$\bar G^r({\bf r}, {\bf r}^\prime; t)=0$ for $t<0$. 

We introduce also another retarded Green's function 
$G^r$ that is related to $\bar G^r$ by an equation
\begin{equation}
\frac{\partial}{\partial {\bf r}}\Theta({\bf %
r})\frac{\partial}{\partial {\bf r}}\bar G^r({\bf r}, {\bf r}^\prime; \omega)%
=\frac{\partial^2 }{\partial {\bf r}^2} G^r({\bf r}, {\bf r}^\prime;
\omega)~.
\label{Greens_relation}
\end{equation}
Taking into account Eqs.\ (\ref{phi_n})-(\ref{Greens_relation}), we 
immediately obtain for $G^r({\bf r}, {\bf r}^\prime; \omega)$
the eigenmode expansion given by Eq.\ (\ref{plasm_grf}).
Dyadic Green's function $G^r_{\alpha\beta}$ and the corresponding
dyadic renormalized potential $W_{\alpha\beta}$ are defined as
\begin{equation}
G^r_{\alpha\beta}(\mathbf r, \mathbf r^\prime;\omega)=
\frac{\partial^2}{\partial r_\alpha\partial r^\prime_\beta}
G^r(\mathbf r, \mathbf r^\prime;\omega)~,~~~
W_{\alpha\beta}(\mathbf r, \mathbf r^\prime;\omega)=
\frac{\partial^2}{\partial r_\alpha\partial r^\prime_\beta}
V(\mathbf r-\mathbf r^\prime)+
\frac{4\pi}{\varepsilon_h}
G^r_{\alpha\beta}(\mathbf r, \mathbf r^\prime;\omega)~,~~~
\label{diadic_Greens}
\end{equation}
where $\alpha,\beta=x,y,z$. 

\subsection{Nonradiative and Radiative Relaxation of NQD in Proximity of
Nanoplasmonic System}

An important process, which can contribute to 
the linewidths $\gamma_a$ and $\gamma_d$ of the NQD donor and acceptor
transitions, is the nonradiative transfer
of energy from NQDs to a metal. The corresponding contribution to  
linewidths $\gamma_m$ can be found in a straightforward way (cf.\ 
Ref.\ \onlinecite{Novotny_Hecht_2006_Principles_of_Nanooptics}) to have the form
\begin{equation}
\gamma_m=-\frac{2\pi|\mathbf d|^2}{3\hbar \varepsilon_h}
\mathrm{Im}\,G^r_{\alpha\alpha}(\mathbf r,\mathbf r;\omega)~,
\label{gamma_m}
\end{equation}
where $\mathbf d$ is either $\mathbf d_d$ or $\mathbf d_a$, and $\mathbf
r$ is the position of the NQD, which is considered as a point-like
object.

Beside the FRET and the energy transfer to the metal (nonradiative
relaxation), there is also a process of the radiative relaxation that
can also be enhanced by the nanoplasmonic system (nanoantenna effect --
see, e.g., Ref.\ \onlinecite{Novotny_Hecht_2006_Principles_of_Nanooptics}).
The corresponding transition renormalized (SP-enhanced) dipole 
$\mathbf d^{(r)}$ and the radiative
relaxation rate $\gamma_r$ are given by
\begin{equation}
d^{(r)}_\alpha=d_\alpha+d_\beta\int\Theta(\mathbf r^\prime)
G_{\alpha\beta}(\mathbf r,\mathbf r^\prime;\omega)d^3 r^\prime~,~~~
\gamma_r=\frac{4\omega^3}{3\hbar c^3}\left|\mathbf d^{(r)}\right|^2~,
\label{gamma_r}
\end{equation}
where $\mathbf d$ is the bare transition dipole.

\subsection{Renormalized Coulomb Interaction for Nanoshells}

For nanoshells, the eigenfunctions
are given by products of the radial power functions and spherical harmonics
that describe the angular dependence. The renormalized Coulomb potential 
$W(\mathbf r,\mathbf r^\prime;\omega)$ for $r,r^\prime>a$, where $a$ is the 
external radius of the nanoshell, acquires the form
\begin{equation}
W(\mathbf r,\mathbf r^\prime;\omega)=
V(\mathbf r-\mathbf r^\prime)+
\frac{4\pi}{\varepsilon_h a}
\sum_{lm}\frac{F_l(x,\omega)}{2l+1}
\left(\frac{a^2}{rr^\prime}\right)^{l+1}
Y_{lm}(\mathbf r) Y_{lm}^\ast(\mathbf r^\prime)~.
\label{W_shell}
\end{equation}
Analogous expressions for the $r$ and/or $r^\prime$ belonging to the inner part 
of the shell have also been obtained (not shown) and used in the computations, 
in particular, those illustrated in Fig.\ \ref{W.eps}.
In Eq.\ (\ref{W_shell}), the spherical 
harmonics $Y_{lm}$ depend only on the directions of the
corresponding vectors, and $x$ is the shell aspect ratio (i.e., the ratio of the 
inner to outer shell radius). Form factor $F_l$ is given by a resonant
pole expression:
\begin{equation}
F_l(x,\omega)=\frac{s^+_l f^+_l}{s(\omega)-s^+_l}+ 
\frac{s^-_l f^-_l}{s(\omega)-s^-_l}~,
\label{F_l}
\end{equation}
where $P=\pm$ refers to symmetry of the corresponding SP mode and
\begin{equation}
f_l^{\pm}=\pm\frac{(\pm\lambda+1)(2l+1\pm\lambda)}{4\lambda(l+1)}~,~~~
s_{l}^{\pm}=\frac{l+(1\mp\lambda)/2}{2l+1}~,~~~
\lambda=\sqrt{1+4l(l+1)x^{2l+1}}~.
\label{f_s_lambda}
\end{equation}
Note that SP eigenmodes with the $P=+$ symmetry have dominating oscillator
strength in the long-wavelength (red and near-ir) part of the spectrum, 
where the quality factor of the SP resonances for noble metals
is high, and which are most important in many cases.

\subsection{FRET and Quenching Rates for NQDs on Nanoshell}

We will calculate here the FRET rate $\bar\gamma_F$ averaged over the position
of the acceptor on the nanoshell. Because we are interested in the
SP-enhanced transfer over the distances much exceeding the usual
F\"orster range, we will disregard the bare Coulomb potential $V(\mathbf
r-\mathbf r^\prime)$. Then substituting the SP eigenfunctions and
eigenvalues into Eq.\ (\ref{gamma_m}) and integrating over the solid
angle of the vector $\mathbf r^\prime$, we obtain
\begin{equation}
\bar\gamma_F=\frac{2\pi\vert\mathbf d_a\vert^2 \vert\mathbf d_d\vert^2}{9\varepsilon_h^2
\hbar^2 a^6} 
J \sum_{l=1}^{\infty}\left|F_l(x,\omega)\right|^2(2l+1)(l+1)^2
\left(\frac{a^2}{r r^\prime}\right)^{2l+4}~.
\label{bar_gamma_F}
\end{equation}

The rate $\gamma_m$ of the excitation quenching due to the
energy transfer to the metal on the nanoshell is obtained from Eq.\
(\ref{gamma_m}) by the substitution of the SP eigenfunctions and
eigenvalues. This procedure is actually greatly simplified, without
affecting the result, if the angular averaging is performed. This leads
to a simple expression 
\begin{equation}
%\begin{aligned}
\gamma_m=-\frac{|\mathbf{d}|^2}{6\varepsilon_h\hbar a^3}
\sum_{l=1}^{\infty}\mathrm{Im}\, F_l(x,\omega)(2l+1)(l+1)
\left(\frac{a}{r}\right)^{2l+4}~.
\label{bar_gamma_m}
\end{equation}

This work was supported by grants from the Chemical Sciences,
Biosciences and Geosciences Division of the Office of Basic Energy
Sciences, Office of Science, U.S. Department of Energy, a grant
CHE-0507147 from NSF, a grant from the US-Israel BSF, and by the DOE
Center for Integrated Nanotechnologies jointly operated by the the Los
Alamos and Sandia National Laboratories.  MIS gratefully
acknowledges useful discussions with D.\ Bergman and A.\ Nitzan.

Correspondence and requests for materials
should be addressed to MIS~(email: mstockman@gsu.edu)

%% Put the bibliography here, most people will use BiBTeX in
%% which case the environment below should be replaced with
%% the \bibliography{} command.

%\bibliography{../../texbib/references}

\end{document}